\documentclass[sigconf, screen, language=english]{acmart}
\usepackage{subcaption}
\usepackage{listings}
\usepackage[normalem]{ulem}
\usepackage{xcolor}
\usepackage{seqsplit}

\definecolor{black}{rgb}{1,1,1}
\definecolor{lightpurple}{RGB}{243, 226, 244}

\AtBeginDocument{%
  }

\copyrightyear{2026}
\acmYear{2026}
\setcopyright{cc}
\setcctype{by}
\acmConference[LAK 2026]{LAK26: 16th International Learning Analytics and Knowledge Conference}{April 27-May 01, 2026}{Bergen, Norway}
\acmBooktitle{LAK26: 16th International Learning Analytics and Knowledge Conference (LAK 2026), April 27-May 01, 2026, Bergen, Norway}

\begin{document}

\title{LLM-based Multimodal Feedback Produces Equivalent Learning and Better Student Perceptions than Educator Feedback}

\author{Chloe Qianhui Zhao}
\email{cqzhao@cmu.edu}
\affiliation{%
  \institution{Carnegie Mellon University}
  \city{Pittsburgh}
  \state{PA}
  \country{USA}
}

\author{Jie Cao}
\email{jiecao@unc.edu}
\affiliation{%
  \institution{The University of North Carolina at Chapel Hill}
  \city{Chapel Hill}
  \state{NC}
  \country{USA}
}

\author{Jionghao Lin}
\email{jionghao@hku.hk}
\authornote{Corresponding Author}
\affiliation{%
  \institution{The University of Hong Kong}
  \streetaddress{Pokfulam Rd}
  \city{Hong Kong}
  \country{China}
}
\affiliation{%
  \institution{Carnegie Mellon University}
  \city{Pittsburgh}
  \state{PA}
  \country{USA}
}

\author{Kenneth R. Koedinger}
\email{koedinger@cmu.edu}
\affiliation{%
  \institution{Carnegie Mellon University}
  \city{Pittsburgh}
  \state{PA}
  \country{USA}
}

\renewcommand{\shortauthors}{Zhao et al.}

\begin{abstract}
Providing timely, targeted, and multimodal feedback helps students quickly correct errors, build deep understanding and stay motivated, yet making it at scale remains a challenge. This study introduces a real-time AI-facilitated multimodal feedback system that integrates structured textual explanations with dynamic multimedia resources, including the retrieved most relevant slide page references and streaming AI audio narration. In an online crowdsourcing experiment, we compared this system against fixed business-as-usual feedback by educators across three dimensions: (1) learning effectiveness, (2) learner engagement, (3) perceived feedback quality and value. Results showed that AI multimodal feedback achieved learning gains equivalent to original educator feedback while significantly outperforming it on perceived clarity, specificity, conciseness, motivation, satisfaction, and reducing cognitive load, with comparable correctness, trust, and acceptance. Process logs revealed distinct engagement patterns: for multiple-choice questions, educator feedback encouraged more submissions; for open-ended questions, AI-facilitated targeted suggestions lowered revision barriers and promoted iterative improvement. These findings highlight the potential of AI multimodal feedback to provide scalable, real-time, and context-aware support that both reduces instructor workload and enhances student experience.
\end{abstract}



\maketitle

\section{Introduction}
Feedback is fundamental to effective learning because it helps students close the gap between their current and desired performance~\cite{hattie_power_2007}. Yet delivering personalized and immediate feedback at scale remains difficult under high student–teacher ratios~\cite{nicol_formative_2006}, and competing instructional responsibilities~\cite{kane_relationships_2019,shute_focus_2008, carless_development_2018}. The challenge intensifies for multimodal feedback (integrating modalities such as text explanations, visual annotations, and audio narration) aligned to course materials because instructors may need to locate appropriate resources and compose them into coherent, context-relevant messages. Although multimodal designs can leverage multiple cognitive channels to improve learning~\cite{mayer_multimedia_2020}, their manual preparation is time-consuming and hard to scale in real classrooms.

Recent advances in Large Language Models (LLMs) with multimodal capabilities create new opportunities for immediate, contextualized feedback at scale~\cite{jin_chatting_2025, kasneci_chatgpt_2023}. When coupled with retrieval of course-specific resources, such systems can provide personalized, context-aware support across self-study, blended learning, and large online courses, offering flexible availability and rapid response. Properly designed, multimodal feedback may enhance retention via dual coding, support diverse accessibility needs, and increase engagement through richer interactions~\cite{paivio_dual_1990, mayer_multimedia_2020}.

However, important gaps remain in understanding the actual implementation of multimodal feedback. First, most prior work focuses on single modality (e.g., text-only feedback)~\cite{ryan_feedback_2019}, leaving open how real-time multimodal compositions affect learning experience. Learners may value AI feedback for availability and consistency but still question trustworthiness or struggle to integrate cross-channel information~\cite{cristea_slideitright_2025}. Poorly coordinated modalities can induce information overload and increase extraneous cognitive load, ultimately hindering learning~\cite{mestre_chapter_2011, cristea_slideitright_2025}. Second, despite rapid model progress, the educational effectiveness of the newest real-time LLMs, including streaming, low-latency variants (e.g., OpenAI Realtime API~\cite{openai_introducing_2025}) and next-generation models (e.g., GPT-5~\cite{openai_gpt-5_2025}), remains empirically under-documented in authentic learning tasks. Their speed and multimodal composition capabilities may change design trade-offs (e.g., text length vs. audio narration vs. visual retrieval), but how these newer models perform in practice is unknown. Addressing these uncertainties is essential before classroom deployment of AI-facilitated multimodal feedback.

Comprehensive evaluation of the learning effect of such systems should include: learning gains, learner engagement (log data analysis), learner perceptions~\cite{keuning_systematic_2018}. Learning gains establish pedagogical effectiveness, whereas engagement captured in log data reflects how learners interact with the feedback in situ~\cite{jovanovic_predictive_2019}; perceptions (e.g., clarity, trust, motivation, usability) shape uptake and self-regulated learning behaviors~\cite{jovanovic_predictive_2019, rudian_feedback_2025}. Large, tightly controlled classroom trials are often impractical due to recruitment limits, instructional disruption, and the deployment overhead of novel systems. Given these constraints, we adopt an online crowdsourcing study to recruit learners online and gather diverse learning data efficiently, maintain experimental control without burdening instructors, and iterate on design prior to in-class implementation. Notably, prior work shows that short-term performance indicators from well-designed online tasks can serve as informative proxies for longer-term learning trajectories~\cite{gao_predicting_2025}, offering actionable signals about system effectiveness ahead of classroom rollout. To address the research gaps, we pose three \textbf{R}esearch \textbf{Q}uestions (\textbf{RQs}):
\begin{quote}
\begin{itemize}
    \item[\textbf{RQ1:}] How effectively does AI-facilitated multimodal feedback support learning?
    \item[\textbf{RQ2:}] How do learners engage with AI-facilitated multimodal feedback?
    \item[\textbf{RQ3:}] How do learners perceive the value and quality of AI-facilitated multimodal feedback?
\end{itemize}
\end{quote}

\begin{figure*}[h]
  \centering
  \includegraphics[width=0.78\textwidth]{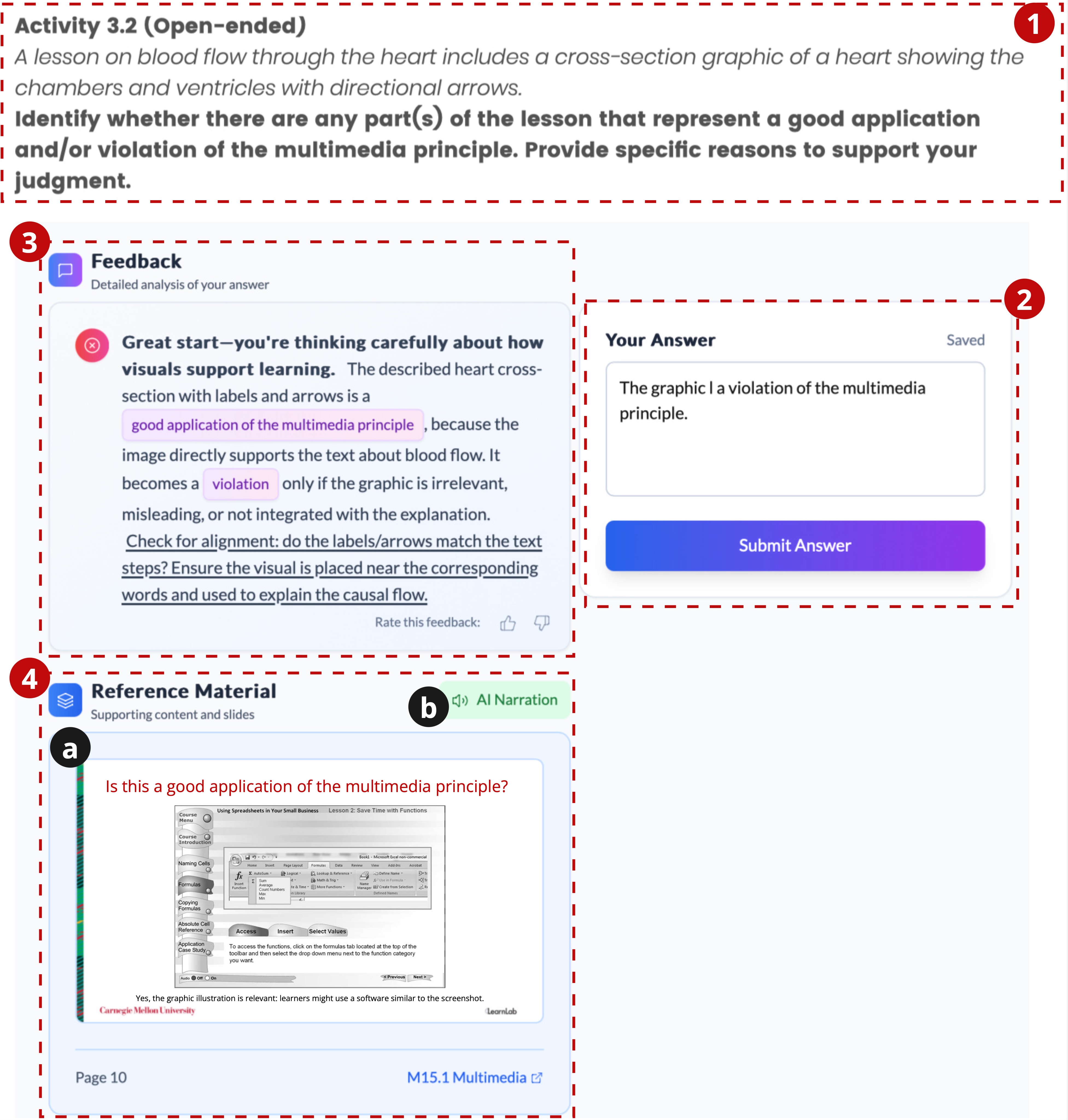}
  \caption{A screenshot of the learner interface of the AI multimodal feedback system
  when embedding for an open-ended learning activity. Above the embedded interface: (1) the practice question. On the right side: (2) learner's answer panel. On the left side: (3) visually enhanced AI corrective feedback that guides learners’ understanding: \textbf{correctness information}, \colorbox{lightpurple}{highlighted key concepts} with their further explanations provided as hover tooltips to keep only text directly relevant to the correction in the main feedback), and \uline{actionable advice}; and (4) reference materials, including (4a) the most relevant slide page with its slide file link, and (4b) a button to activate AI audio narration for the slide page.
  }
  \label{fig:interface}
  \Description[Annotated learner interface showing question prompt, answer input, feedback panel, and reference material sections.]{The figure shows a feedback system interface divided into clearly labeled regions marked with numeric and letter annotations.
At the top, a horizontally spanning region labeled “1” contains the question text. It includes a bold heading followed by instructional text describing a learning task about blood flow through the heart and asking learners to judge whether elements of the lesson reflect a good application or a violation of the multimedia principle.
Below this region, the interface is split into two vertical panels. On the right, labeled “2,” is an answer input area titled “Your Answer.” A single-line text field contains a learner response stating that the graphic is a violation of the multimedia principle. A submit button is positioned below the input field.
On the left, labeled “3,” is a feedback panel titled “Feedback.” The panel displays written feedback in paragraph form. The first sentence is visually emphasized in bold. Within the following sentence, two key phrases are highlighted to contrast a good application of the multimedia principle with a violation. A subsequent sentence is visually distinguished with underlining to emphasize guidance about aligning labels, arrows, and explanatory text.
Below the feedback panel, a section labeled “4” is titled “Reference Material.” Inside this section, a slide preview marked “a” displays a screenshot of spreadsheet software with menus and toolbars. The slide includes a question at the top and footer text indicating a page number and a linked lesson title. To the right of the section heading, a circular control marked “b” is labeled “AI Narration,” indicating an option to play audio narration.}
\end{figure*}

In this paper, we present two main contributions:
\begin{enumerate}
    \item
    A real-time AI-facilitated multimodal feedback system (shown in Fig.~\ref{fig:interface}) that integrates structured text explanations with relevant multimedia resources.
    \item a randomized controlled online crowdsourcing experiment comparing AI-facilitated multimodal feedback
     with a baseline of fixed business-as-usual feedback created by educators across three dimensions: learning gain, learner engagement (usage logs), and learner perception.
\end{enumerate}

\section{Related Work}
\subsection{Feedback in Digital Learning Environments}
Feedback plays a central role in learning by helping students understand how to close the gap between their current and desired performance~\cite{nicol_formative_2006}. Classic theories emphasize three qualities of effective feedback: timeliness, specificity, and actionability~\cite{hattie_power_2007,shute_focus_2008, ryan_designing_2021}. Timeliness ensures that feedback arrives while cognitive processing of the task is still active, enabling immediate correction and reinforcement~\cite{hattie_power_2007}; specificity provides clear, criterion-referenced guidance rather than vague evaluations, reducing uncertainty and guesswork~\cite{goodman_feedback_2004}; actionability goes beyond diagnosis to offer concrete strategies for improvement, increasing the likelihood of meaningful learner action~\cite{ryan_designing_2021}.

Despite its importance, providing high-quality, individualized feedback at scale remains challenging, especially in large-enrollment classes and massive open online courses~\cite{barros_large_2025, gabbay_combining_2024}. Along the human pathway, educator-authored feedback is rich and nuanced but constrained by time and workload, limiting scalability for targeted responses~\cite{nicol_formative_2006, aldino_analytics_2025}; peer or crowdsourced feedback can redistribute evaluative effort, yet often suffers from inconsistency and variable quality~\cite{carless_development_2018,shin_comparing_2025}. To address these constraints, work has progressed in automated feedback generation. Earlier automated feedback often relied on comparing student responses with a predefined desired answer~\cite{cavalcanti_automatic_2021}, or utilized techniques like Natural Language Processing (NLP) and rule-based approaches~\cite{park_students_2024}. These systems were applied in educational tasks, particularly structured tasks, such as programming scoring in STEM field~\cite{keuning_systematic_2018} and Automatic Writing Evaluation (AWE) for language learning~\cite{rudian_feedback_2025}. However, they generally required more domain expertise input and were hard to transfer to new settings or domains without extensive retraining~\cite{jurgensmeier_generative_2024}.

More recently, researchers have explored large language models (LLMs) to generate feedback~\cite{shin_comparing_2025, cao_first_2025, venugopalan_combining_2025}. LLM-based feedback can be adaptive and context-aware, alleviating manual authoring burdens and improving responsiveness~\cite{kasneci_chatgpt_2023}. For instance, LLMs have shown the potential to scale educationally effective feedback through a learner-centered design approach~\cite{cao_first_2025}. Yet the majority of studies to date still focus on text-only outputs, underutilizing the multimedia resources of modern digital learning environments (e.g., images, audios, videos, and other digital course materials)~\cite{lin_mufin_2024, aldino_analytics_2025}. This narrow focus limits support for diverse learner needs and overlooks opportunities to leverage multiple representational channels for clarity and accessibility~\cite{lin_mufin_2024, mayer_multimedia_2020}.

\subsection{Multimodal Feedback: Theories and Practices}
The transition from text-only to multimodal feedback represents a major development in educational technology. Building on Mayer’s multimedia learning theory, which posits separate visual and auditory channels for processing information~\cite{mayer_multimedia_2020}, multimodal feedback that combines text, visuals, and audio can, in principle, support diverse learners, enable dual coding, and address accessibility needs~\cite{paivio_dual_1990}. At the same time, because each channel has limited capacity, integrating multiple modalities raises cognitive-load concerns~\cite{lee_multimodality_2025, cristea_slideitright_2025}. Cognitive load theory holds that working memory is constrained; when multimodal elements are poorly coordinated, learners may experience split-attention, redundancy, and other forms of extraneous load that impede learning~\cite{mestre_chapter_2011}.
Effective multimodal design therefore hinges on three considerations~\cite{mestre_chapter_2011,mayer_multimedia_2020}: (i) relevance—selecting a minimal set of complementary modalities rather than overwhelming learners with overloaded information; (ii) contiguity—ensuring spatial and temporal alignment between text, visuals, and audio to reduce integration effort; and (iii) learner control—offering optional, progressively disclosed detail so learners can adjust depth and pace. These tensions complicate large-scale, timely, targeted feedback: producing high-quality multimodal feedback typically requires substantial manual effort to create, align, and synchronize resources~\cite{cristea_slideitright_2025, aldino_analytics_2025}. As a result, many educational systems continue to rely on simpler, single-modality approaches despite the theoretical appeal of multimodal feedback~\cite{ryan_feedback_2019}.


\subsection{AI-Facilitated Multimodal Feedback}
Recent advances in Generative Artificial Intelligence (GenAI) have enabled real-time, context-sensitive multimodal feedback. Contemporary large language models (LLMs) such as GPT-5~\cite{openai_gpt-5_2025}, Claude Sonnet 4.5~\cite{anthropic_claude_2025}, and Gemini 2.5 Pro~\cite{gemini_team_gemini_2025}, are capable of processing multimodal inputs (such as text, images, and audios) and producing integrated multimodal outputs. For instance, existing research explored creating communicative AI tutors with natural tones~\cite{chen_vtutor_2025}, and processing visual information in course materials~\cite{cristea_slideitright_2025}. Through techniques such as retrieval-augmented generation (RAG)~\cite{lewis_retrieval-augmented_2021}, these models can dynamically retrieve and integrate course-specific resources such as lecture slides, multimedia materials, and other supplementary documents directly into the feedback to provide richer and relevant context and streamline the selective information presentation process~\cite{cristea_slideitright_2025, jin_chatting_2025}.

Despite these advances, research on AI multimodal feedback is still at an early stage~\cite{lin_mufin_2024}. First, most prior work mainly examined the learning effect of one modality beyond text, such as audio feedback\cite{wang_t2a-feedback_2025} and graphic visualizations~\cite{cavalcanti_automatic_2021}, without combining multiple modalities as a coherent whole. Second, the capabilities of cutting-edge technology such as GPT Realtime API and GPT-5, which promise lower latency and richer integration of diverse modalities, have rarely been examined in educational contexts~\cite{sesler_towards_2025}. Third,  without careful coordination, multimodal outputs risk introducing cognitive challenges such as split-attention effects, redundancy, and extraneous cognitive load when learners must reconcile multiple, potentially conflicting streams of information~\cite{mestre_chapter_2011, mayer_multimedia_2020, mestre_chapter_2011}. These issues highlight the importance of involving pedagogical theories and deliberate system design~\cite{sesler_towards_2025}, including synchronizing modalities, clarifying the role of each information component, and limiting feedback to the most relevant and actionable elements~\cite{mayer_multimedia_2020}.

Overall, while generative AI opens unprecedented opportunities for delivering real-time multimodal feedback, rigorous empirical research is still needed to evaluate its effectiveness and to understand its impact on learning experiences for scalable deployment in educational settings.

\section{The design rationale of AI multimodal feedback system}
\subsection{Learner Interface Design and Integrability}

In this study, we developed an AI-facilitated multimodal feedback system with
a learner-facing interface designed for the integration into existing online intelligent tutoring systems (ITSs). As shown in Fig.~\ref{fig:interface}, the feedback interface is embedded directly beneath a practice item via an \texttt{<iframe>} (an HTML element that allows an external web page or application to be displayed within another page). This allows learners to interact with 
the feedback without leaving the host platform, preserving task focus~\cite{mestre_chapter_2011} while extending the ITS with real-time, AI-facilitated multimodal feedback.

The interface is divided into two main parts: The right side provides input fields for learners to submit initial responses and to revise answers, supporting both multiple-choice questions (MCQs) and open-ended questions (OEQs). The left side dynamically displays feedback provided by the system. The feedback design follows learner-centered principles emphasizing clarity, accessibility, and cognitive load reduction~\cite{ryan_designing_2021, hattie_power_2007}. By placing feedback adjacent to the input area, learners can immediately compare their responses with system guidance, enabling rapid iteration and sustained engagement~\cite{mayer_multimedia_2020}.

\subsection{Multimodal Inputs}
Our AI multimodal feedback system accepts multimodal inputs to construct contextually appropriate feedback. Inputs include (1) the original question (text and, when applicable, images) designed by educators and used in normal educational practice, (2) the learner’s answer (i.e., MCQ option selection or OEQ response), (3) the researcher-designed prompt that instructs AI feedback generation, and (4) a knowledge base derived from lecture slides, where relevant slide pages are represented using the visual understanding capabilities of \texttt{\seqsplit{gpt-5}}.

\subsection{Multimodal Feedback Composition}
The multimodal output generated by
the feedback system comprises two coordinated components: (1) visually enhanced corrective feedback and (2) a resource area that combines a relevant slide page with audio narration. These components are designed to complement each other and are grounded in existing studies on multimedia learning and feedback.

\subsubsection{Corrective Feedback}
Text remains the primary modality to preserve clarity and interpretability. The AI-facilitated corrective feedback follows a learner-centered framework~\cite{ryan_designing_2021} to provide an organized, actionable summary of performance, including: (1) correctness evaluation, (2) explanations, and (3) actionable suggestions for improvement. The prompt used to generate this feedback is provided in the GitHub repository: \url{https://github.com/zqh0421/llm-based-feedback-lak26}.

\subsubsection{Resource Area: Visual and Auditory Supplements}
Below the structured textual feedback, the system presents a resource area that leverages additional modalities to enhance comprehension and engagement.
\begin{itemize}
    \item The most relevant slide page is dynamically retrieved and displayed to visually ground the feedback in course-specific materials.
    \item The AI-generated narration explains the slide page while connecting the information of the practice question, learner's answer, and the textual feedback above. When learners experience fatigue from reading slide text or need alternative forms of explanation, they can activate the audio narration feature.
\end{itemize}

This dual-channel presentation is informed by Mayer’s multimedia learning theory~\cite{mayer_multimedia_2020} and Paivio’s dual-coding theory~\cite{paivio_dual_1990}, which together suggest that coordinated use of verbal and non-verbal channels can enhance information processing and retention. Importantly, the narration does not duplicate the text verbatim; instead, it provides complementary explanations with natural tones to keep learners engaged.

\subsection{Structured Textual Feedback Generation for Visual Enhancement}
To address the design challenge of balancing clarity, pedagogy, and cognitive load~\cite{cristea_slideitright_2025}, 
our feedback system follows a structured pipeline for generating visually enhanced textual feedback. When a learner submits a response, model \texttt{\seqsplit{gpt-5}} (\texttt{\seqsplit{gpt-5-2025-08-07}}) is prompted to produce concise feedback and configured with \texttt{\seqsplit{verbosity=low}} to reduce information redundancy.

The generated feedback is returned as a JSON object following the schema below:

\begin{minipage}{\linewidth}
\begin{lstlisting}
{
    "score": [0 for incorrect, 1 for correct],
    "structured_feedback": "<statement>[Supportive assessment of their attempt].</statement> <explanation>[Clear, encouraging explanation with <term explanation='[helpful context]'>[key concepts]</term>].</explanation><advice>[Constructive suggestions including reflective questions].</advice>"
}
\end{lstlisting}
\end{minipage}

This structured output includes:
\begin{itemize}
    \item A numeric score indicating correctness: for MCQs, \verb|0| (incorrect) or \verb|1| (correct); for OEQs, \verb|0| (incorrect), \verb|1| (partially correct), or \verb|2| (correct). This score is mapped to a color-coded indicator for visual cues.
    \item An HTML fragment whose tags support later visual formatting in the frontend: \verb|<statement>|, \verb|<explanation>|, and \verb|<advice>| for the main feedback elements.
    \item Inline \verb|<term>| tags for context-aware keyword tooltips that are not directly related to the corrective information. These appear on demand to keep the main feedback concise and to reduce the cognitive load.
\end{itemize}

Once processed, the feedback is formatted and rendered within the left feedback panel, as shown in Fig.~\ref{fig:interface} (3). This integration of visual cues with structured text directly supports feedback principles of timeliness, specificity, and actionability~\cite{hattie_power_2007}, while minimizing extraneous cognitive load~\cite{mayer_multimedia_2020}.


\subsection{RAG-Based Reference Retrieval}
To ensure that reference resources are contextually relevant,
the system employs a feedback generation workflow using retrieval-augmented generation (RAG)~\cite{gao_retrieval-augmented_2024}. Model \texttt{\seqsplit{gpt-5}}'s vision understanding information and its semantic vector from \texttt{\seqsplit{text-embedding-3-small}} are pregenerated for each slide page and practice question and saved in the database. For each question, the system retrieves the top 3 most relevant slide pages by cosine similarity to the question vector. When a learner submits a response, the best-match slide page is presented in the resource area, and the top-3 slide descriptors are provided to the model, together with the question content and learner answer, to generate feedback grounded in course materials.

\subsection{Feedback Timeliness Enhancement}
Timely feedback allows learners to act while their cognitive processing of the task remains active~\cite{hattie_power_2007}. By combining strategies including optimized model configurations, cached slide embeddings,  streaming narration, and pre-generated MCQ feedback, 
the system delivers timely multimodal feedback and improve learning experience.
\begin{itemize}
\item We configure \texttt{\seqsplit{gpt-5}} (\texttt{\seqsplit{gpt-5-2025-08-07}}) with \texttt{\seqsplit{reasoning\_effort=low}} as a deliberate trade-off: modestly reduced depth for substantially lower latency, while preserving essential accuracy.
\item The \texttt{\seqsplit{gpt-5}} vision understanding information and its semantic vector for each slide page and practice question are pre-generated and saved in the database, to avoid producing too much visual information during learner's interaction.
\item AI Audio narration is streamed via the \texttt{\seqsplit{gpt-realtime}} API, allowing playback to begin before the entire narration is generated. This design helps minimize the delay of audio feedback.
\item In addition, because MCQ answers are drawn from a limited option set, the system pre-generates feedback for each option when practice questions are uploaded, reducing wait time at interaction.
\end{itemize}

\section{Experimental Design}
\subsection{Recruitment}
We conducted an online crowdsourcing study by recruiting participants through Prolific\footnote{\url{https://www.prolific.com/}}. Eligibility criteria included being at least 18 years old, currently enrolled as a university or college student, and physically located in the United States during the study. Participants were required to have sufficient English proficiency to fully understand the consent form, learning materials, tasks, and feedback. To ensure full system functionality, they were required to use a laptop or desktop computer with a reliable internet connection. Participants received \$14.00 USD for approximately 60 minutes of participation, contingent on completing the required study tasks. This compensation rate followed Prolific’s ethical payment guidelines and met the platform’s recommendation for an ``excellent'' pay level to maintain participant motivation. Before beginning the study, participants provided informed consent electronically via Qualtrics\footnote{\url{https://www.qualtrics.com/}}. They were explicitly informed that they could withdraw at any time without penalty, and that their data would be de-identified to protect confidentiality. The study protocol was reviewed and approved by the Institutional Review Board (IRB) at Carnegie Mellon University.

\subsection{Learning Materials}
The learning materials for this experiment were drawn from normal educational practice in a learning science course and were used in both online and in-person settings. The content focuses on applying the \emph{multimedia principle} in e-learning design, with two learning goals: (1) Describe the multimedia principle and explain why it supports learning and (2) Recognize when the principle has been violated and when it has been applied effectively.


\subsection{Procedures}
The entire study was hosted on the Qualtrics survey platform and completed by participants in a single session. Instead of an ITS, we chose Qualtrics as our experimental host because it enables efficient collection of learner data and management of experimental flow in our controlled study, while also supporting the same web-based \texttt{<iframe>} embedding mechanisms commonly available in ITSs and learning management systems. The procedure consisted of four stages: (1) pre-test, (2) learning-by-doing activities, (3) post-test, and (4) post-survey. Attention-check questions were embedded throughout the study to ensure data quality, and only participants who passed these checks were included in the final analyses.

\begin{itemize}
    \item \textbf{Pre-Test.}
    The pre-test consisted of 14 multiple choice questions (MCQs) and 2 open-ended questions (OEQs). This is to establish participants' baseline knowledge before any learning or feedback exposure. No detailed feedback was provided to avoid influencing baseline performance; only the final total score for the MCQs was displayed at the end of this stage.
    \item \textbf{Learning-by-Doing Activities.} Participants then completed 13 learning-by-doing activities (8 MCQs and 5 OEQs) designed to practice and reinforce the learning objectives through active engagement. After submitting each response, participants received immediate feedback that varied depending on their assigned experimental condition. Learner's usage log data were captured during this stage, including time spent on tasks, and learning behaviors such as submission count and answer revisions.
    \item \textbf{Post-Test.} After completing the practice phase, participants were presented with a post-test consisting of the same set of questions as the pre-test. Using identical questions in both pre- and post-tests allowed for direct within-subject comparisons, ensuring that differences in performance were due to learning effects rather than changes in test difficulty.
    \item \textbf{Post-Survey.} Finally, participants completed a post-survey to reflect on their experience and perceptions of the feedback they received. The survey included Likert-scale items measuring their perceived quality and value of the provided feedback.
\end{itemize}

\subsection{Conditions}
The difference between conditions was the type of feedback provided during the Learning-by-Doing phase. Note that all participants, regardless of condition, were given access to the same lecture slide deck via a web link. The slide file consisted of text and images aligned with the learning goals and contained an embedded short instructional video with audio narration recorded by the course instructor (voice only, no face) as optional supplementary learning material. A sample slide page from this deck is shown in Fig.~\ref{fig:interface} (4a). All other aspects of the study, including learning materials, study tasks, and interface, were also identical to ensure a consistent user experience. As shown in Fig.~\ref{fig:comparison}, only the information shown in the left panel of 
the learner interface varied by condition:

\begin{figure*}[t!]

  \centering
  \begin{minipage}{0.8\textwidth}
  \begin{subfigure}[b]{0.4\linewidth}
    \centering
    \includegraphics[width=\linewidth]{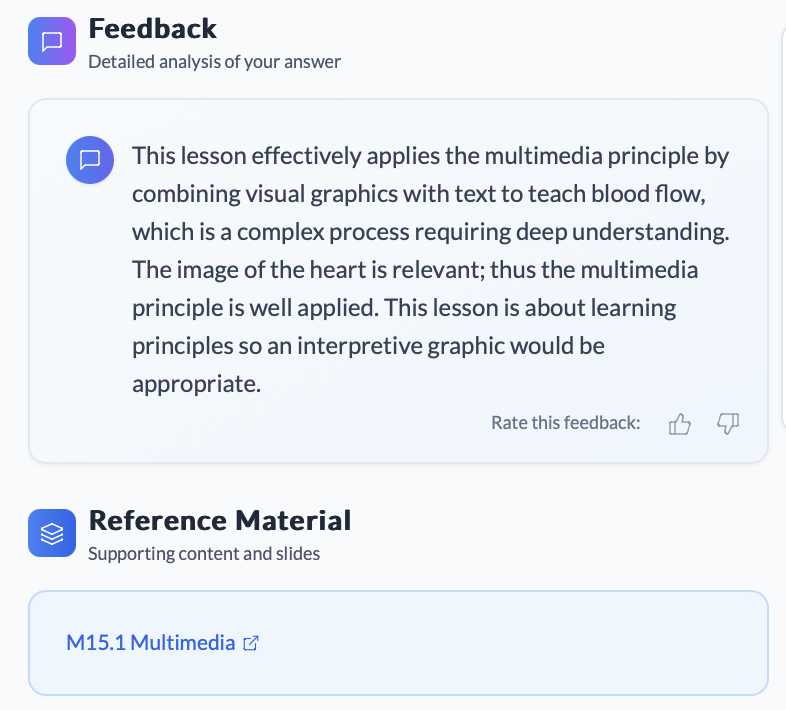}
    \caption{Business-as-usual feedback}
  \end{subfigure}\hfill
  \begin{subfigure}[b]{0.37\linewidth}
    \centering
    \includegraphics[width=\linewidth]{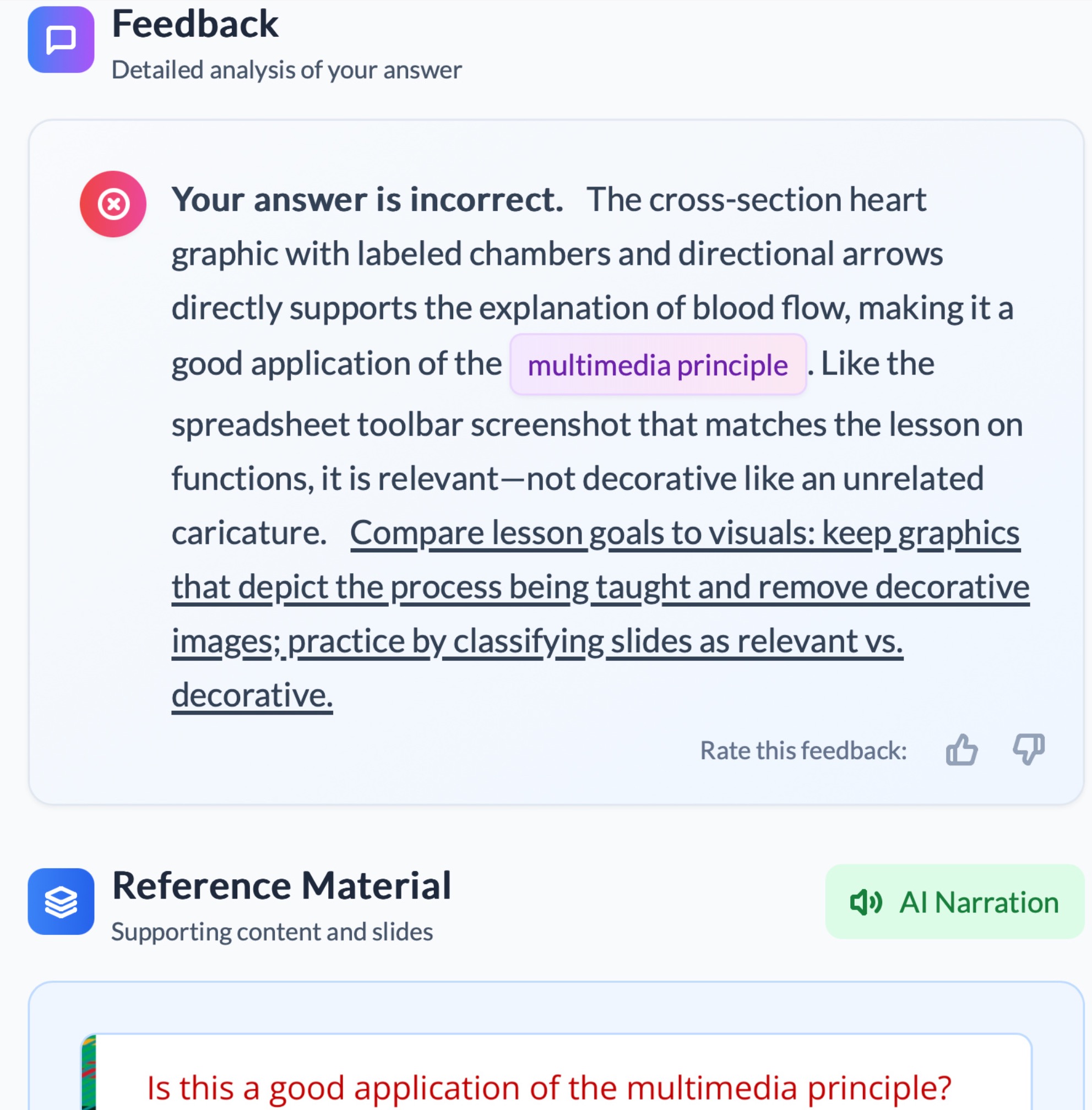}
    \caption{AI Multimodal feedback}
  \end{subfigure}
  \end{minipage}
  \caption{Comparison of the feedback composition in different conditions.}
  \label{fig:comparison}
\end{figure*}

\begin{itemize}
    \item \textbf{Business-as-usual Feedback (Baseline).} Learners received fixed, pre-written text feedback authored by experienced educators. Slide resources were provided via a single link to the full slide deck, without highlighting which slide page was most relevant to a specific learning task. This set of feedback (i.e., general text feedback and a lecture slide file) corresponded to materials used in normal educational practice. Repeated submissions to a certain question redisplayed the same feedback.
    \item \textbf{AI Multimodal Feedback (Treatment).} Learners received dynamically generated, structured feedback (Fig.~\ref{fig:interface}), including visual-enhanced AI corrective feedback, automatically retrieved most-relevant slide page, and optional audio narration.
\end{itemize}

Upon providing consent, participants were assigned to one of the conditions and informed about the feedback source (AI or human) and the available modalities before engaging in the learning activities.

\section{Results}
\subsection{Participants}
After attention-check screening during learning stage, we retained data from participants who were at least 18 years old, currently located in the United States and enrolled as a student in a university or college, and pre-screened on Prolific. The final sample contains 197 learners (52.8\% female, 37.1\% male, 3.5\% others, and 6.6\% chose not to disclose) from diverse major backgrounds (such as science, engineering, law and humanities), with an average age of $32.5 \pm 9.07$ years. They were randomly assigned to the two conditions: 87 in Business-as-usual Feedback group and 110 in AI Multimodal Feedback group.

\begin{figure}[b]
  \centering
  \includegraphics[width=0.83\linewidth]{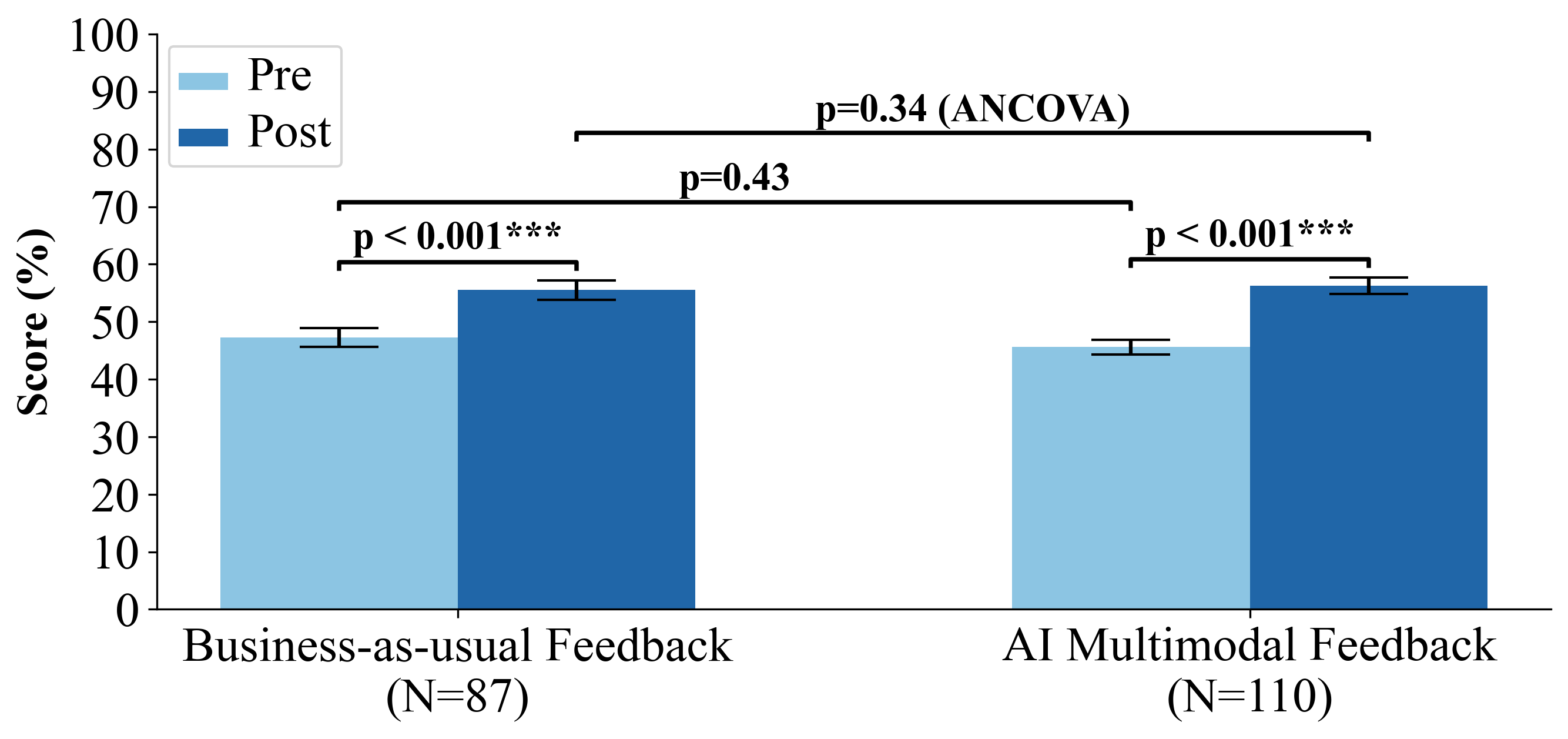}
  \caption{Pre- and Post-Test Scores (\%) Showing Comparable Significant Learning Gains in both Baseline and AI Multimodal Feedback Conditions. Error bars represent ±SEM (Standard Error of the Mean).}
  \label{fig:pre-post}
  \Description[Grouped bar chart showing pre-test and post-test scores across two feedback conditions.]{The figure is a grouped bar chart comparing test scores across two experimental conditions. The vertical axis is labeled “Score (percent)” and ranges from 0 to 100.
Two groups of bars are shown along the horizontal axis. The left group represents the Business-as-usual Feedback condition with a sample size of 87. The right group represents the AI Multimodal Feedback condition with a sample size of 110. Within each group, two adjacent bars correspond to pre-test and post-test scores.
In both groups, the post-test bar is higher than the pre-test bar. Each bar includes an error bar representing the standard error of the mean. Brackets above each group indicate statistically significant within-group differences between pre-test and post-test scores, both labeled with p-values smaller than 0.001.
Additional two brackets span across groups to indicate between-group statistical comparisons. One bracket above the pre-test bars is labeled with a p-value of 0.43. A higher bracket spanning the post-test bars is labeled with an ANCOVA result of p = 0.34.}
\end{figure}

In this study, 87 responses for the Business-as-usual Feedback group and 110 responses for the AI Multimodal Feedback group were collected. Each learner response includes valid pre-test, post-test, and learning-by-doing logs. The two OEQs were scored by a zero-shot AI grader using a detailed rubric. To ensure rating validity for open-ended question scoring, 10\% of the learner responses were coded by both expert human grader and AI grader to calculate the inter-rater reliability. The agreement was $\kappa=0.862$ for OEQ1 and $\kappa=0.944$ for OEQ2. The pre- and post-test scores for both groups were standardized to a percentage scale (0–100\%), as shown in Fig.~\ref{fig:pre-post}.

\subsection{RQ1.1: Learning Evaluation}

A t-test was conducted on the pre-test scores to examine whether the two groups were comparable before the learning stage. The results revealed no significant difference between groups ($t = 0.9534$, $p = 0.3417$, $Cohen's\ d = 0.1395$), which indicates that participants in both groups began the study with similar baseline knowledge levels.

Within-group analyses were then conducted to determine whether each learning condition led to significant improvements in participants' performance from pre-test to post-test. For the Business-as-usual Feedback group, a paired-samples t-test showed that post-test scores were significantly higher than pre-test scores ($t = -3.4841$, $p = 0.0006$***, $Cohen's\ d = 0.5376$), indicating that original human-provided feedback had a meaningful and positive impact on learning outcomes. The AI Multimodal Feedback group also demonstrated significant improvement from pre- to post-test ($t = -5.6273$, $p < 0.001$***, $Cohen's\ d = 0.7623$).

We then conducted an ANCOVA with post-test score as the dependent variable, condition as a between-group factor, and pre-test score as a covariate. The assumption of homogeneity of regression slopes was satisfied ($p = 0.24$). After controlling for pre-test performance, the effect of condition on post-test scores was not significant ($F = 0.91$, $p = 0.34$, partial $\eta^2 = 0.0047$).

Both Business-as-usual Feedback and AI Multimodal Feedback were effective in significantly improving learners' knowledge levels, as evidenced by the substantial within-group gains observed in each condition. AI-Multimodal Feedback yields outcomes comparable to those of original educator feedback according to the between-group comparison.

\subsection{RQ1.2: Time Efficiency}
We calculated the system response time for each question type, which includes the total time required to return both the structured text feedback (including score, tooltip, statement/advice tags, etc.) and the slide reference. For multiple choice questions, the median of system response time is 0.299 seconds ($Mean = 0.700$, $SD = 1.457$); for open-ended questions, the median of system response time is 6.23 seconds ($Mean = 7.210$, $SD = 3.389$).

A limitation of this study is that the human feedback was authored before the research as part of routine instruction; consequently, we did not record the time educators spent writing it. As a proxy, we compared word counts between the LLM prompt used to generate feedback and the corpus of human-authored feedback used in this study. For AI feedback, the human-authored input to the LLM (excluding the retrieval of relevant course materials) consisted of the question, its options (and the correct option), the student’s answer, and the prompt instructions for MCQs, and the question, student’s answer, and prompt instructions for open-ended questions. The prompt template itself contained 356 words, whereas the human text feedback totaled 985 words across 8 multiple-choice questions (26 options in total) and 5 open-ended questions. Assuming comparable effort per word for preparing a set of reusable feedback, the system is approximately $985/356 \approx 2.77\times$ more time-efficient than authoring equivalent human text feedback. In addition to that, educators would require additional time to tailor feedback to individual student responses for open-ended questions, link multimedia resources, or provide visual/audio feedback to match the system’s functionality. Taken together, the system enables students to receive richer, more targeted feedback more quickly while reducing educator workload.

\subsection{RQ2: Learner Engagement}
We evaluate learner engagement through two main metrics.

\subsubsection{Total learning time.} Participants in the Business-as-usual Feedback group spent a median of 21.13 minutes on learning-by-doing activities, whereas those in the AI Multimodal Feedback group spent a median of 25.44 minutes. An independent-samples \textit{t}-test indicated no significant difference in learning time between groups ($t=-0.413$, $p=0.680$, $Cohen's\ d=-0.062$).

\subsubsection{Number of active-learning actions.} Active learning refers to students’ deliberate engagement in tasks rather than passively receiving information~\cite{prince_does_2004}. In this study, we quantified and recorded active learning as the total count of submit/update actions used to either to request new feedback or to update their own answers (i.e., clicks to submit or resubmit an answer).
A Mann–Whitney U test was conducted (Table~\ref{tab:action-comparison}) and showed no significant overall difference between Original Educator Feedback and AI Multimodal Feedback ($U=4075.5$, $p=0.0738$), indicating comparable aggregate levels of active-learning behavior across conditions. By question type, however, patterns diverged: for multiple-choice items, participants in the baseline condition submitted more attempts than those in the AI condition ($U = 5660.5$, $p = 0.02$*), consistent with greater trial-and-error under human feedback; for open-ended items, the AI condition showed a trend toward more submissions ($U = 4060.5$, p = 0.066).

\begin{table*}[h]
\centering
\footnotesize
\caption{Between-group Comparison of the Number of Action-Learning Actions in the Learning-by-Doing Stage (Business-as-usual Feedback vs.\ AI Multimodal Feedback)}
\label{tab:action-comparison}
\begin{tabular}{lccll}
\toprule
& \multicolumn{2}{c}{\textbf{Mean (SD)}} & & \\
\cmidrule(lr){2-3}
\textbf{Question Type} & \multicolumn{1}{c}{Business-as-usual Feedback} & \multicolumn{1}{c}{AI Feedback} & \multicolumn{1}{c}{\textbf{$U$}} & \multicolumn{1}{c}{\textbf{$p$}} \\
\midrule
Overall                  & 21.39 (9.05) & 21.94 (5.78) & 4075.5 & 0.074 \\
Multiple-Choice Questions& 13.78 (7.28) & 11.92 (2.96) & 5660.5 & \textbf{0.020*} \\
Open-Ended Questions     &  7.61 (4.12) &  8.44 (3.35) & 4060.5 & 0.066 \\
\bottomrule
\end{tabular}
\end{table*}

\subsection{RQ3: Learner Perception}
A total of 81 valid post-survey responses were retained for the Business-as-usual Multimodal Feedback group and 99 for the AI Multimodal Feedback group after filtering post-survey responses based on completion and attention check questions. Table~\ref{tab:survey_results} presents the Mann–Whitney U significance testing results for three main survey dimensions: overall perception, perceived quality, and perceived value of the feedback they had received.

\subsubsection{Overall Perception.}
Participants reported significantly lower mental effort when interacting with the AI Multimodal Feedback ($U = 5190.50$, $p = 0.0006$***), while their satisfaction scores were also significantly higher for AI feedback compared to original educator feedback ($U = 3310.50$, $p = 0.0240$*).

\subsubsection{Perceived Quality.}
AI feedback received significantly higher ratings for specificity, clarity, and simplicity (all $p < 0.05$), indicating that participants found the AI-facilitated multimodal feedback to be clearer, more understandable, more concise, and more focused than original educator feedback. No significant difference was observed for correctness ($p = 0.5378$), suggesting that learners perceived the factual accuracy of AI and original educator feedback to be similar.

\subsubsection{Perceived Value.}
AI feedback scored significantly higher for motivation ($U = 3174.50$, $p = 0.0146$*), suggesting that participants felt more encouraged to engage with the task after receiving AI-generated feedback. Meanwhile, no significant differences were observed for trust ($p = 0.1456$) and acceptance ($p = 0.2393$). This indicates that the AI multimodal feedback system achieved a comparable level of perceived trustworthiness and acceptance as original educator feedback, which contrasts with prior studies where trust in AI feedback was reported to be much lower~\cite{cristea_slideitright_2025}.

Taken together, these results demonstrate that the AI feedback system not only met educator-level expectations in trust, acceptance and correctness but also surpassed Business-as-usual feedback in multiple dimensions, including overall satisfaction, clarity and understandability, simplicity, correctness,
and motivational aspects (all $p < 0.05$). Notably, no significant differences were observed for \textit{Correctness}, \textit{Trust}, and \textit{Acceptance}
($p \geq 0.05$). This indicates that the AI feedback system has reached a comparable level of trustworthiness and acceptance as original educator feedback.

\begin{table*}[t]
\centering
\small
\caption{Between-group comparison of Survey Likert Results (Business-as-usual Feedback vs.\ AI Multimodal Feedback)}
\label{tab:survey_results}
\setlength{\tabcolsep}{6pt}
\begin{tabular}{llccll}
\toprule
& & \multicolumn{2}{c}{\textbf{Mean (SD)}} &   & \\
\cmidrule(lr){3-4}
\textbf{Dimension} & \textbf{Sub-dimension} & Business-as-usual Feedback & AI Multimodal Feedback & \multicolumn{1}{c}{$U$} & \multicolumn{1}{c}{$p$} \\
\midrule
Overall Perception
  & Mental Effort & 7.22 (1.63) & 6.17 (2.09) & 5190.50 & \textbf{0.0006***} \\
  & Satisfaction  & 4.19 (1.00) & 4.46 (0.90) & 3310.50 & \textbf{0.0240*} \\
\midrule
Perceived Quality
  & Specificity   & 3.78 (0.91) & 4.28 (0.75) & 2638.00 & \textbf{0.0001***} \\
  & Clarity       & 4.48 (0.76) & 4.68 (0.67) & 3346.50 & \textbf{0.0199*} \\
  & Simplicity    & 3.75 (1.02) & 4.09 (0.93) & 3246.00 & \textbf{0.0207*} \\
  & Correctness   & 4.00 (0.97) & 4.02 (1.12) & 3806.50 & 0.5378 \\
\midrule
Perceived Value
  & Motivation    & 3.70 (1.06) & 4.04 (1.01) & 3174.50 & \textbf{0.0146*} \\
  & Trust         & 4.11 (0.85) & 4.24 (0.93) & 3543.00 & 0.1456 \\
  & Acceptance    & 4.23 (0.73) & 4.33 (0.75) & 3614.00 & 0.2393 \\
\bottomrule
\end{tabular}

\vspace{2pt}
\parbox{\linewidth}{\footnotesize\raggedright
\textit{Note.} Other than mental effort, which is rated from 1 (low) to 10 (high), all other items are rated on a 5-point Likert scale. The full questionnaire is available at the GitHub repository: \url{https://github.com/zqh0421/llm-based-feedback-lak26}.
}
\end{table*}

\section{Discussion}
\subsection{LLM-based multimodal feedback produces learning equivalent to original educator feedback (RQ1).}

The AI multimodal feedback yielded learning gains that were not significantly different from those produced by original human-crafted feedback, consistent with prior comparisons between human and AI-based feedback for supporting student learning~\cite{alsaiari_emotionally_2025,cristea_slideitright_2025}.  While average gains were slightly higher and the effect size larger for the AI condition, both conditions showed significant improvement from pre-test to post-test. This indicates that well-designed AI feedback can match the instructional quality of traditional educator-created feedback. In addition, considering the time efficiency for both instructors and learners, the real-time, structured, and context-aware AI multimodal feedback provides immediate and actionable guidance that could produce positive effect on students' learning. Such immediacy has the potential to help instructors reduce feedback-related workload and allow students to receive timely multimodal feedback that meaningfully supports their learning.

Our results suggest that AI multimodal feedback can be a scalable complement to human instruction, particularly in environments where quick turnaround and individualized feedback are critical. Moreover, because the feedback system could be easily embedded as \texttt{\seqsplit{<iframe>}} component, it can be integrated into existing digital learning platforms, without requiring fundamental changes to course design. This compatibility reduces adoption barriers and supports gradual integration into real-world classrooms. However, due to the limited session time and constrained task scope, our study cannot address long-term effects or performance on delayed post-tests. Further experiments are needed in real classroom settings, ideally combining in-person observations and online log data, to evaluate lasting learning outcomes and secure broader instructor support for adoption.

\subsection{LLM-based feedback shifted effort: fewer MCQ retries, more OEQ revisions, same time on task (RQ2).}

Learning-by-Doing log data showed that overall active-learning actions (e.g., total submissions and answer revisions) were similar across both groups. However, differences emerged by question type: participants in the baseline condition submitted more frequently on MCQs, while those in the AI condition submitted more frequently on OEQs. No significant differences were observed in total time-on-task between groups. For MCQs, the static educator-provided feedback used in this study often focused on option-level explanations, encouraging learners to engage in a form of trial-and-error exploration. In contrast, the AI feedback more directly articulated why a chosen option was incorrect and how to reason toward the correct answer. This more specific guidance might have helped learners identify more efficiently toward the target concept, which aligns with prior work suggesting that concrete, actionable feedback can limit unproductive trial-and-error behavior~\cite{goodman_feedback_2004}. For OEQs, AI-generated structured suggestions provided targeted and actionable guidance aligned with learners' specific answers. This lowered the barrier for revising open-ended responses, which is reflected in the observed trend toward a higher number of OEQ submissions in the AI condition.

These findings highlight the importance of tailoring feedback strategies by question type. For MCQs, systems should discourage superficial trial-and-error behaviors and focus on guiding conceptual understanding. For OEQs, structured, targeted scaffolding can foster deeper engagement and higher-quality revisions. Learning analytics should also consider multimodal indicators like revision quality to better capture meaningful engagement.

\subsection{LLM-based feedback improved clarity, specificity, conciseness, motivation, and satisfaction (RQ3).}
Across multiple subjective dimensions, the AI multimodal feedback was rated significantly higher than the Business-as-usual feedback for clarity, specificity, conciseness, motivation, and overall satisfaction. Importantly, ratings for correctness, trust, and acceptance were statistically comparable between conditions, where learners were explicitly informed whether the feedback were from human or AI. Together, these findings suggest that the AI multimodal feedback enhanced the overall learning experience, which may help sustain continued learner motivation~\cite{alsaiari_emotionally_2025}.

Participants also reported significantly lower mental effort when using the AI feedback, suggesting that the multimodal design alleviated perceived cognitive load in our study, where the main goal was to reduce extraneous load so that learners could focus their limited resources on high-value processes of integrating and understanding information relevant to their learning tasks. This contrasts with prior work in which learners experienced increased cognitive load when receiving AI feedback, as they had difficulty efficiently locating essential information due to the overly lengthy responses without pointing out key points~\cite{cristea_slideitright_2025}. These results indicate that the AI system was able to deliver feedback that learners perceived as both effective and usable. The combination of visual structuring, context-relevant multimodal resources, and layered information helped learners quickly process and act on feedback, aligning with best practices in feedback design. The findings provide strong evidence for the design principle of structured, progressively disclosed, and context-enhanced feedback. This design approach can be adopted in future online learning platforms to improve both the perceived quality and usability of feedback without compromising accuracy or trust.

\section{Conclusion}
In a controlled online crowdsourcing setting, this study compared fixed Business-as-usual feedback provided by educators with real-time multimodal AI feedback. Results show that the AI approach achieves learning outcomes equivalent to human feedback while significantly outperforming it on perceived clarity, specificity, conciseness, motivation, overall satisfaction, and lower cognitive load, thereby elevating perceived quality and value with comparable correctness, trust, or acceptance. Coupled with learning process log evidence of distinct behavioral patterns, this profile supports a design strategy that enables on-demand multimodal augmentation—scalable for broad deployment, configurable across diverse learning contexts, and flexible to embed within existing intelligent tutoring systems (ITSs). The proposed combination of real-time, structured, and context-aware multimodal feedback offers a practical path to reduce instructor workload while providing students with actionable, timely, and targeted support that yields measurable learning gains.

\section{Limitations and Future Work}

While this study provides evidence of the potential benefits of real-time multimodal AI feedback, several limitations remain to be acknowledged, along with directions for future research.

The experiment was conducted in a controlled online crowdsourcing setting with a single one-hour session. This design choice was made as a trade-off to further support motivation and data quality by limiting the number of questions to reduce time pressure associated with Prolific’s maximum session duration~\cite{cristea_slideitright_2025}. Nevertheless, it restricted our understanding of long-term learning outcomes and transfer effects. Future work should extend to authentic classroom environments with longitudinal data collection, examining delayed testing, knowledge retention, and sustained engagement, and should combine online log data with in-class observations to provide a more holistic view of how learners interact with feedback in real-world contexts and how AI multimodal feedback may influence learners’ deep learning processes~\cite{stadler_cognitive_2024}.

In addition, our participant pool consisted of U.S.-based university and college students recruited via Prolific, which, while relevant for many digital learning platforms, limits generalizability. Further studies should therefore include more diverse learner groups, subject domains, and cultural contexts to better understand how multimodal feedback functions across broader scenarios.

Third, the observed differences between the Business-as-usual feedback condition and the AI feedback condition may be attributable to multiple confounded factors, such as the presentation style of slides, different feedback source (AI vs. human), and the current study cannot disentangle their individual contributions to learning outcomes. In the prior 2×2 study, we did not observe statistically significant differences in learning gains across isolated feedback dimensions~\cite{cristea_slideitright_2025}, which motivated the present study’s deliberate adoption of a focused gambling strategy, which prioritized testing an theoreticaly most promising condition rather than exhaustive variable isolation~\cite{bruner_toward_1966}. Nevertheless, future work should employ orthogonal manipulations of feedback design factors to disentangle their individual and interactive effects and examine the impact of individual feedback components and their interactions.

Lastly, we were only able to measure overall perceived load, which does not allow us to disentangle different types of cognitive load. Therefore, we cannot distinguish reductions in extraneous load from potential changes in generative load. While lower extraneous load is generally desirable, increased generative load can sometimes be beneficial, as it may reflect productive struggle and deeper cognitive engagement. Future work should employ more fine-grained measures to better capture how AI feedback influences different cognitive processes during learning.

Addressing these limitations will enable future research to build on our findings and develop scalable, adaptive feedback systems that integrate into classrooms, reduce instructor workload, and provide learners with personalized support.

\begin{acks}
This research was supported by the Generative AI + Education Tools R\&D Seed Grant at Carnegie Mellon University and by a grant from the URC (Grant No. 2401102970) at the University of Hong Kong. The opinions, findings, and conclusions expressed in this paper are solely those of the authors. We are grateful to all participants, anonymous reviewers, our lab members and colleagues for their insightful feedback and support during this study.
\end{acks}

\bibliographystyle{ACM-Reference-Format}
\bibliography{references}










\end{document}